\documentclass[10pt,twocolumn,letterpaper]{article}

\usepackage[pagenumbers]{cvpr} 

\usepackage{graphicx}
\usepackage{amsmath}
\usepackage{amssymb}
\usepackage{booktabs}
\usepackage{algorithm}
\usepackage{algorithmicx}
\usepackage{algpseudocode}
\usepackage{multirow}
\usepackage[table,xcdraw]{xcolor}

\usepackage[pagebackref,breaklinks,colorlinks]{hyperref}

\usepackage{booktabs}

\usepackage[capitalize]{cleveref}
\crefname{section}{Sec.}{Secs.}
\Crefname{section}{Section}{Sections}
\Crefname{table}{Table}{Tables}
\crefname{table}{Tab.}{Tabs.}

\def\cvprPaperID{****} 
\def\confName{CVPR}
\def\confYear{2022}

\begin{document}
\title{A Heterogeneous Graph Learning Model for Cyber-Attack Detection }

\def\eg{e.g\onedot}
\def\etc{etc\onedot}
\definecolor{redcol}{rgb}{1, 0, 0}
\newcommand{\red}[1]{\textcolor{redcol}{#1}} 

\newcommand*{\affaddr}[1]{#1} 
\newcommand*{\affmark}[1][*]{\textsuperscript{#1}}
\newcommand*{\email}[1]{\texttt{#1}}
\author{Mingqi Lv, Chengyu Dong, Tieming Chen, Tiantian Zhu, Qijie Song, Yuan Fan}

\maketitle


\begin{abstract}
   A cyber-attack is a malicious attempt by experienced hackers to breach the target information system. Usually, the cyber-attacks are characterized as hybrid TTPs (Tactics, Techniques, and Procedures) and long-term adversarial behaviors, making the traditional intrusion detection methods ineffective. Most existing cyber-attack detection systems are implemented based on manually designed rules by referring to domain knowledge (e.g., threat models, threat intelligences). However, this process is lack of intelligence and generalization ability. Aiming at this limitation, this paper proposes an intelligent cyber-attack detection method based on provenance data. To effective and efficient detect cyber-attacks from a huge number of system events in the provenance data, we firstly model the provenance data by a heterogeneous graph to capture the rich context information of each system entities (e.g., process, file, socket, etc.), and learns a semantic vector representation for each system entity. Then, we perform online cyber-attack detection by sampling a small and compact local graph from the heterogeneous graph, and classifying the key system entities as malicious or benign. We conducted a series of experiments on two provenance datasets with real cyber-attacks. The experiment results show that the proposed method outperforms other learning based detection models, and has competitive performance against state-of-the-art rule based cyber-attack detection systems.
\end{abstract}

\section{Introduction}
\label{sec:intro}
Nowadays, cyber-attacks are usually launched by experienced hackers, and characterized as hybrid TTPs (Tactics, Techniques, and Procedures) and long-term adversarial behaviors. Thus, it is very difficult to detect them by utilizing traditional intrusion detection methods\cite{2018A,2020Preparing}, which usually looks at short-term behaviors by ignoring the rich context information of the system.

Aiming at this problem, we investigate the experiences of many existing researches and find that provenance data are the best approach for the cyber-attack detection task\cite{2018Provenance,2017Applying,2020UNICORN} . Provenance data describe system entities (e.g., processes, files, sockets) and interactions between the system entities (the interactions are usually represented as system events, e.g., read, write, fork). First, provenance data can describe how system entities interact with each other and provide rich context information for each system entities\cite{2015Trustworthy}. Second, provenance data can connect causally-related events even when those events are separated by a long period of time\cite{2020UNICORN}. Provenance data are usually collected by using OS audit logs (e.g., Linux Auditd\cite{LINUXAuditd}, Windows ETW\cite{ETW}).

To carry out the cyber-attack detection task by utilizing provenance data, most existing works model the provenance data as a directed acyclic graph (called a provenance graph), where the nodes represent system entities and the edges represent system events. Figure 1 shows an example of the provenance graph. Then, most existing studies (e.g., SLEUTH\cite{2018SLEUTH}, HOLMES\cite{2019HOLMES}, CONAN\cite{0CONAN}) try to detect cyber-attacks by manually designing a variety of rules based on threat knowledge (e.g., kill chain\cite{2015Technical}, ATT\&CK\cite{ATT&CK}). Although the rule based detection strategies are efficient and have achieved state-of-the-art performance, they still have the following limitations. First, the design of rules is an extremely difficult task, since it requires expertized domain knowledge of cyber-attack strategies, operation systems, and computer network. Second, rules lack generalization ability, since they are designed under specific macro environment (e.g., the cyber-attack strategies, the network environment). They easily become obsolete and have to be adjusted or even redesigned, when the macro environment changes. Third, rules are limited to human experiences and cannot handle complex and latent patterns.

On the other hand, learning based techniques have long been applied for cyber security tasks such as intrusion detection\cite{2018A,2020Preparing}, malware detection\cite{2017Malware,2020Enhancing}, and fraud detection\cite{2018Online,2020Interleaved}. As compared with the rule based techniques, learning based techniques have the following advantages. First, learning based techniques (especially deep learning techniques) could extract latent patterns and build detection model from the training dataset with less domain knowledge. Second, the detection model could adapt to new macro environment by retraining or fine-tuning in an automatic manner. Since provenance data are represented as provenance graphs, we exploit GNN (Graph Neural Network)\cite{2019Comprehensive} to build the cyber-attack detection model. However, due to the complexity of cyber-attacks and system provenance, building cyber-attack detection model based on GNN is still a challenging task.
\begin{figure}
    \centering
    \includegraphics[width=0.9\columnwidth]{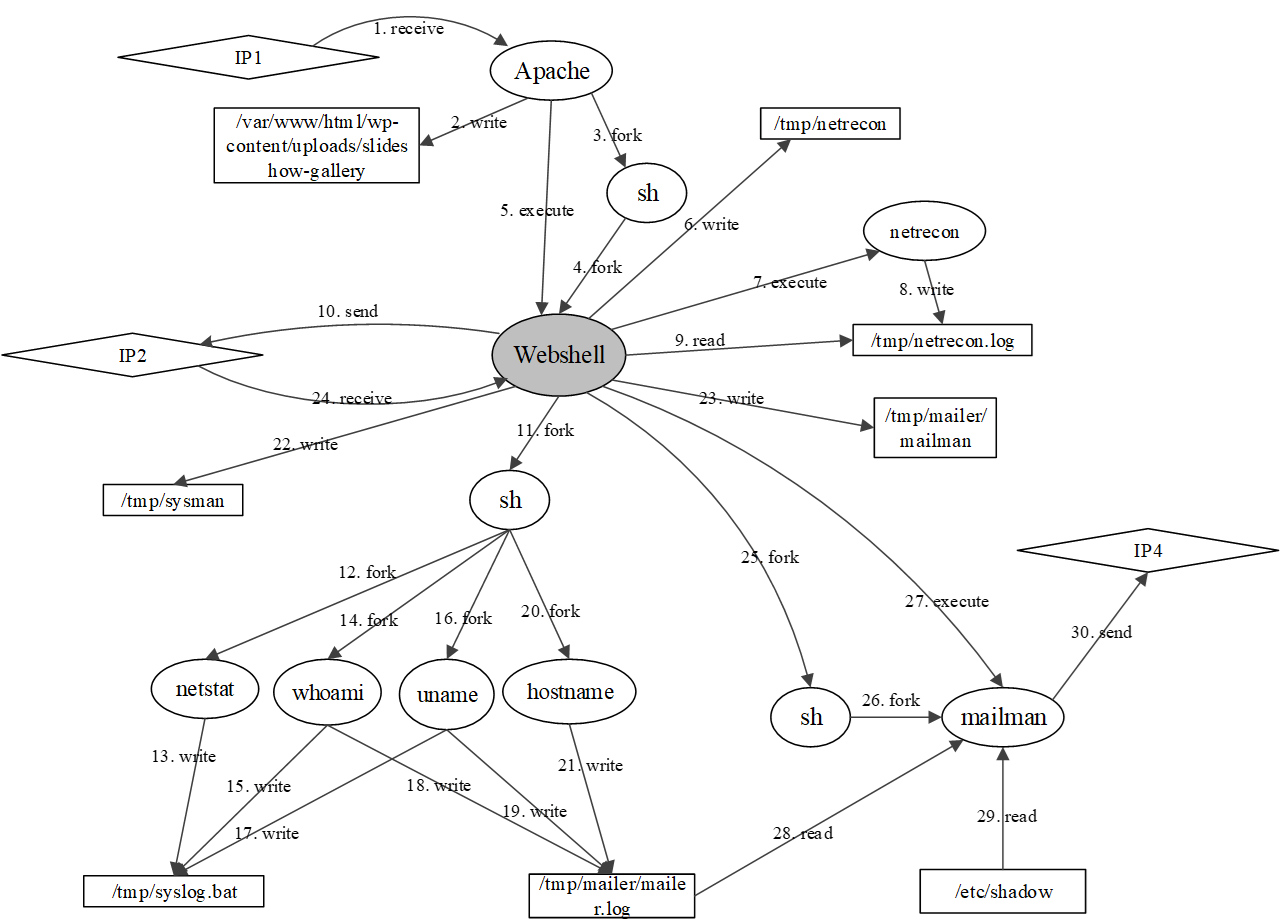}
    \caption{An example of provenance graph.}
    \label{fig:one}
\end{figure}
First, provenance graphs generated from OS audit logs are complex. A provenance graph involves multiple types of nodes (e.g., processes, files, sockets) and multiple types of edges (e.g., read, write, and create). In addition, a node or an edge can be depicted by additional attributes (e.g., a file can be depicted by attributes such as source, file extension, and sensitive level).

Second, the nodes and edges of provenance graphs are specific instances of system entities and system events, which make the detection model almost impossible to learn common patterns from these totally distinct instances.

Third, cyber-attack detection is a speed sensitive task. However, most GNNs have high computation and space cost. In addition, since cyber-attacks are usually persistent, the provenance graphs continue to grow over time, making the GNN based methods inevitably suffer from efficiency and memory problems.

Fourth, cyber-attacks are usually stealthy, so it is very difficult to detect them by only looking at individual system entities (e.g., an attack activity might be accomplished by the collaboration of multiple processes and files). However, it is also impractical to consider the whole provenance graph, which is usually too large and dominated by benign system entities.

Fifth, the detection model has to continuously deal with new incoming system entities, while GNNs learn the representations of new system entities by embedding them with the existing nodes and edges in a shared space. However, it is infeasible to rerun the learning procedures whenever new system entities arrive due to the high computation complexity.

Aiming at the above challenges, this paper proposes an intelligent cyber-attack detection system based on heterogeneous provenance graph learning. For the first and second challenges, we adopt a heterogeneous graph\cite{0Mining} to represent the complex system entities and events in the provenance graph, and then uses meta-path based approach\cite{2017metapath2vec} to extract higher-level semantic interactions among the system entities. For the third challenge, we use graph embedding technique to learn a low-dimensional vector representation for each node in the heterogeneous graph. Here, the graph embedding can be executed as a pre-training step, and the vector representations are then used to quickly initialize the detection model. For the fourth and fifth challenges, we link new incoming system entities to the existing heterogeneous graph and perform a local graph sampling to incorporate the new incoming system entities and the most related existing nodes into a small and compact local graph. Then, the detection model is built based on this local graph.
In summary, the main contributions of this paper are as follows.

1) We propose an intelligent cyber-attack detection system by leveraging GNNs. It can automatically learn patterns from the provenance data and use for cyber-attack detection without domain knowledge.

2) We propose a heterogeneous graph based approach to model provenance data and learn semantic vector representations for various system entities.

3) We propose a local graph sampling based approach to perform cyber-attack detection upon new incoming system entities, which do not have pre-learnt vector representations.

4) We conducted extensive experiments based on datasets collected by simulating a variety of cyber-attacks on real hosts. The experiment results show that our method outperforms existing machine learning based cyber-attack detection methods and can achieve competitive performance as compared with state-of-the-art rule-based cyber-attack detection systems built on in-depth domain knowledge.

\section{Related Work}\label{sec:related}
As the mainstream of the existing cyber-attack detection systems, forensic analysis approaches aim to discover the attack events and attack paths based on predefined rules. For example, PrioTracker\cite{2018Towards} enables timely cyber-attack analysis by prioritizing the investigation of abnormal causal dependencies, in which the priority of a system event is measured by predefined rules. SLEUTH\cite{2018SLEUTH} identifies system entities and events that are most likely involved in cyber-attacks based on a tag-based approach. Specifically, it firstly designs tags to encode the trustworthiness and sensitivity of code and data, and then it designs rules for cyber-attack detection by leveraging these tags. HOLMES\cite{2019HOLMES} is a hierarchical framework for cyber-attack detection. The key component of HOLMES is an intermediate layer that maps low-level audit data to suspicious behaviors based on rules from domain knowledge (e.g., ATT\&CK model). CONAN\cite{0CONAN} is a state-based framework. Each system entity (i.e., process or file) is represented in an FSA (Finite State Automata) like structure, where the states are inferred through predefined rules. Then, the sequences of states are used for cyber-attack detection and reconstruction. Thanks to the carefully designed rules, forensic analysis approaches could be effective, efficient, and easy to deploy. However, designing effective rules heavily relies on in-depth inter-discipline domain knowledge, and the rules easily become obsolete if the macro environment changes.

To address the limitations of forensic analysis approaches, researchers tried to build cyber-attack detection model in an automatic way by leveraging learning based techniques (including machine learning and deep learning techniques). For example, Barre et al.\cite{barre2019mining} build a classifier to detect cyber-attacks on top of a set of features (e.g., total quantity of data written, number of system files used) extracted from the provenance data. Berrada et al.\cite{berrada2020baseline} extract Boolean-valued features (called contexts) from the provenance graph, and treat cyber-attack detection as an anomaly detection task by using unsupervised learning technique. Xiang et al.\cite{xiang2020detecting} extract different features from two separated platforms (i.e., PC platforms and mobile platforms), and use several machine learning algorithms to detect cyber-attacks based on the combined features. Zimba et al.\cite{zimba2020modeling} identify hosts exhibiting suspicious malicious activities through a semi-supervised learning framework, which extracts features based on a network clustering algorithm and trains the detection model by using both the labeled and unlabeled data. The above studies detect cyber-attacks based on traditional machine learning techniques, in which the core step is feature extraction. However, feature extraction could be viewed as a simplified version of rule design, and it also relies heavily on domain knowledge.

Recently, deep learning\cite{lecun2015deep} has shown its superior capability of discovering underlying features from big data in a totally automatic way. Thus, a few studies have tried to use deep learning techniques for cyber-attack detection. Since persistence is one of the most fundamental characteristic of cyber-attacks, the collected system events usually span over a long period of time and can be represented as a time series. Therefore, RNNs (Recurrent Neural Networks) such as LSTM (Long Short-Term Memory) and GRU (Gated Recurrent Unit) are mostly applied deep learning models\cite{du2017deeplog,shen2018tiresias,eke2019use}. However, RNNs can only capture the sequential relationships among system events, and ignore other important context information (e.g., different types of interactions among system entities). Aiming at this problem, Liu et al.\cite{liu2019log2vec} proposes log2vec, a cyber-attack detection model based on heterogeneous graph embedding. Specifically, it firstly converts log entries into a heterogeneous graph based on a set of predefined rules. Next, it represents each log entry as a low-dimensional vector using heterogeneous graph embedding technique. Finally, it separates malicious and benign log entries based on the log entry vectors. However, log2vec still has the following limitations. First, the heterogeneous graph is constructed based on predefined rules, and thus log2vec still has a strong rely on domain knowledge. Second, log2vec works in an offline manner, and it has not considered the new incoming system entities.

\makeatletter
\renewcommand{\maketag@@@}[1]{\hbox{\m@th\normalsize\normalfont#1}}%
\makeatother
\section{Preliminary}\label{sec:preliminary}
\subsection{Threat Model}

According to the threat models (e.g., kill chain model\cite{2015Technical}, ATT\&CK model\cite{ATT&CK}), the attackers may conduct cyber-attacks through several stages and use a variety of techniques during each stage. Among these stages, previous studies have found the following invariant parts of the cyber-attacks\cite{0CONAN}. First, the attackers must deploy their code to victims. Second, the final targets of cyber-attacks are stealing sensitive information or causing damage. Third, the attackers will communicate with C\&C server. These invariant parts make possible the detection of new cyber-attacks, and the goal of our method is to detect attacks at any of the invariant stages.

The provenance data (e.g., system events such as file operations and network access) can be collected from kernel level information by using OS audit modules (e.g., Auditd\cite{LINUXAuditd}, ETW\cite{ETW}), which have the advantages of high availability, high credibility, and high stability. In this paper, we focus on analytic capabilities and assume the correctness of the kernel, the provenance data, and the analysis model.

\subsection{System Architecture}
\begin{figure}
    \centering
    \includegraphics[width=0.9\columnwidth]{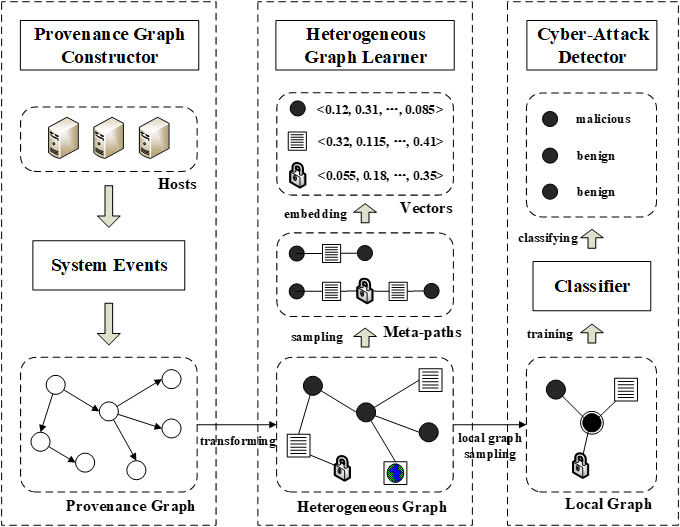}
    \caption{An example of provenance graph.}
    \label{fig:two}
\end{figure}

The architecture of our method is shown in Figure~\ref{fig:two}, consisting of the following three modules.

1) Provenance graph constructor: It continuously collects kernel level system events by using OS audit logs. Then, it arranges these collected system events as a provenance graph, which represents each system event by an edge in the provenance graph.

2) Heterogeneous graph learner: It transforms the provenance graph to a heterogeneous graph, which is compatible with the GNNs. Then, it learns a low-dimensional vector representation for each node (a.k.a. heterogeneous graph embedding), which can preserve the semantical and structural relations between various types of nodes. Specifically, the heterogeneous graph embedding task is performed based on a meta-path sampling strategy and a hierarchical attention based embedding technique.

3) Cyber-attack detector: For a new incoming system entity $v_n$, it links $v_n$ to the existing heterogeneous graph and samples a local small graph by taking $v_n$ as the central node. Then, it treats the cyber-attack detection as a node classification task. In order to discover causal and contextual relationships between system entities by exploring larger graph neighborhoods, it adopts a graph convolutional network for this task.

\section{Methodology}
\label{sec:methodology}

In this section, we first introduce the threat model and the overall architecture of our proposed APT detection system. We then discuss several important topics in its design, including data collection, data compaction, and APT detection framework.

\subsection{Heterogeneous Graph Construction}

First, the provenance data logged by OS audit modules are transformed into a provenance graph. As the example shown in Figure \ref{fig:one}, the nodes represent specific instances of different types of system entities. For example, “mailman” is an instance of process and “/etc/shadow” is an instance of file. Second, we transform the provenance graph into a heterogeneous graph defined as follows.

\noindent \textbf{Definition 1 (Heterogeneous graph).}
A heterogeneous graph, denoted as $G$ = ($V$,$E$), consists of a node set $V$ representing the system entities and an edge set $E$ representing the system events. A heterogeneous graph is also associated with a node type mapping $\phi$: $V\rightarrow A$ and an edge type mapping $\psi$: $E\rightarrow R$. A and R denote the node type set and the edge type set, where $\lvert A\rvert+\lvert R\rvert >2$. It indicates that a heterogeneous graph contains multiple types of system entities and multiple types of system events. The schema [19] of a heterogeneous graph G, denoted as $T_G$ = ($A$, $R$), is a meta-graph with nodes as node types from $A$ and edges as edge types from $R$.

\begin{figure}
    \centering
    \includegraphics[width=0.9\columnwidth]{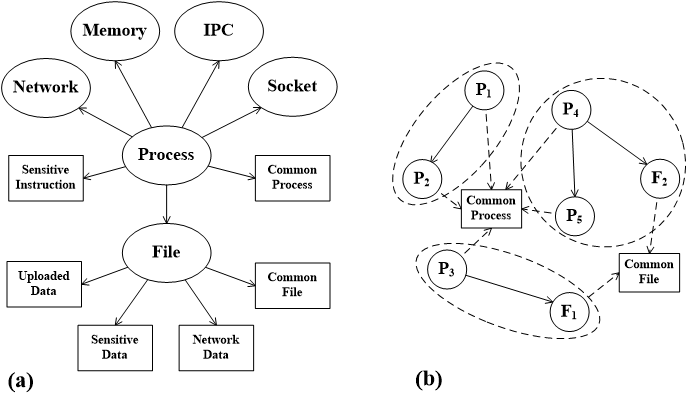}
    \caption{An illustration of the heterogeneous graph: (a) the schema of the heterogeneous graph; (b) an example of the necessity of the “common process” and “common file” attributes.}
    \label{fig:three}
\end{figure}

Based on the definition, the heterogeneous graph is constructed as follows. First, we define the schema of the heterogeneous graph, i.e., the node type set and the edge type set. For our case, we define seven major node types, i.e., process, file, socket, IPC, memory, network, and attribute, to represent system entities. The node type “attribute” is used to describe the features of “process” and “file”, and can be further categorized into six subtypes, i.e., common process, sensitive instruction, common file, network data, sensitive data, and uploaded data. Figure \ref{fig:three}(a) shows these node types and their potential relations. Here, “sensitive instruction” represents a process executes a sensitive instruction, “network data” represents a file contains data from the network, “sensitive data” represents a file contains sensitive data, and “uploaded data” represents a file sent to a remote server. In addition, “common process” and “common file” are redundant attributes, which are used to prevent the heterogeneous graph from being separated into many disconnected small graphs. As the example shown in Figure \ref{fig:three}(b), $P$1, $P$2, $P$3, $P$4, and $P$5 are common files, $F$1 and $F$2 are common files. Without “common process” and “common file” attributes (i.e., the dashed arrows in Figure \ref{fig:three}(b)), these system entities would be separated into three disconnected small graphs (as shown by the dashed ovals), and thus the semantic relatedness between disconnected system entities (e.g., $P$1 and $P$3, $F$1 and $F$2) could not be learnt. This problem could be extremely serious in practice, since the vast majority of system entities are common and benign.

Second, we define eight edge types, as summarized in Table \ref{tab:one}. The edge types represent the relations between system entities (i.e., system events and entity attributes). The intrinsic and complex relations between system entities can provide crucial information for detecting cyber-attacks. Note that a pair of system entity types could have multiple types of relations. For example, a “process” and a “file” could have the relations such as “read”, “write”, “create”, etc.

The main difference between the provenance graph and the heterogeneous graph is as follows. The provenance graph considers the specific instances of system entities, which would make each node totally different and thus hamper the model from learning common patterns from the context of each node. On contrary, the heterogeneous graph uses generalized concepts to define each node. For example, all the file instances share the same node type “file”.

\subsection{Heterogeneous Graph Embedding}
Due to the high persistence of cyber-attacks, the heterogeneous graph could be extremely large. Hence, we should represent and encode the heterogeneous graph in a scalable way to make real-time detection possible. Based on the existing studies, graph embedding is an efficient and effective way to represent large-scale graphs\cite{wang2018billion}, and thus we adopt heterogeneous graph embedding technique to represent the heterogeneous graph. We define heterogeneous graph embedding as follows.
\begin{table*}[]
\caption{A summary of the edge types.}
\label{tab:one}
\resizebox{\textwidth}{!}{
\begin{tabular}{llll}
\hline
\multicolumn{1}{c}{\textbf{Pairs of system entity types}} & \multicolumn{1}{c}{\textbf{ID}} & \multicolumn{1}{c}{\textbf{System event type}} & \multicolumn{1}{c}{\textbf{Description}}                                                         \\ \hline
Process$\rightarrow$ Process              & $R$1 & Process$\rightarrow$op1$\rightarrow$Process       & Here, “op1” can be “fork”, “execute”, “exit”, “clone”, or “change”.                                                                      \\
Process$\rightarrow$ File                 & $R$2 & Process$\rightarrow$op2$\rightarrow$File          & \begin{tabular}[c]{@{}l@{}}Here, “op2” can be “read”, “open”, “close”, “write”, “loadlib”, “create”, “unlink”,\\  “modify”, “truncate”, “rename”, “mmap”, or “update”.\end{tabular} \\
Process$\rightarrow$ Network              & $R$3 & Process$\rightarrow$op3$\rightarrow$Network       & Here, “op3” can be “connect”, “send”, “recv”, “read”, “close”, “accept”, or “write”.                                                     \\
Process$\rightarrow$ Memory               & $R$4 & Process$\rightarrow$op4$\rightarrow$Memory        & Here, “op4” can be “mprotect” or “mmap”.                                                                                                 \\
Process$\rightarrow$ IPC                  & $R$5 & Process$\rightarrow$op5$\rightarrow$IPC           & Here, “op5” can be “write”, “close”,   “read”, or “mmap”.                                                                                \\
Process$\rightarrow$ Socket               & $R$6 & Process$\rightarrow$op6$\rightarrow$Socket        & Here, “op6” can be “connect”, “send”,   “recv”, “read”, “close”, “accept” or “write”.                                                    \\
Process$\rightarrow$ Attribute            & $R$7 & Process$\rightarrow$contain$\rightarrow$Attribute & A process contains one of the following   attributes, i.e., “common process” and “sensitive   instruction”.                              \\
File$\rightarrow$ Attribute               & $R$8 & File$\rightarrow$contain$\rightarrow$Attribute    &  \begin{tabular}[c]{@{}l@{}}A file contains one of the following   attributes, i.e., “common file”, “network data”, \\ “sensitive data”, and “uploaded data”.\end{tabular}  \\\hline        
\end{tabular}}
\end{table*}
\noindent \textbf{Definition 2 (Heterogeneous graph embedding).}
 Given a heterogeneous graph $G$ = ($V$, $E$), the embedding of G is to learn a function $f$: $V \rightarrow R^d$ that maps each node $v \in V$ to a $d$-dimensional vector ($d \ll \vert V \vert$). The low-dimensional vector space should be capable of preserving the topological structure and semantic relations in $G$.
 
Since heterogeneous graphs contain multiple types of nodes and edges, conventional homogeneous graph embedding techniques (e.g., DeepWalk\cite{perozzi2014deepwalk}, node2vec\cite{grover2016node2vec}, LINE\cite{tang2015line}) cannot be directly applied. To address this problem, meta-path was proposed to guide the random walkers to connect various types of nodes\cite{2017metapath2vec}. The random walks on a meta-path can generate a set of neighbors for each node, which can reveal diverse structural and semantic information in a heterogeneous graph. Here, we give the formal definition of meta-path as follows.

\noindent \textbf{Definition 3 (Meta-path).}
A meta-path is a path defined in the schema of a heterogeneous graph, and is in the form of $A_1 \rightarrow R_1\rightarrow A_2\rightarrow R_2\rightarrow …\rightarrow R_{L-1}\rightarrow A_L$, where $R$ = $R_1 \circ R_2 \circ R_{L-1}$   defines a composite relations between node type $A_1$ and $A_L$. In practice, $A_1$ and $A_L$ in a meta-path are usually of the same node type, so that the random walk ends on a meta-path could immediately starts on another one.
Based on the schema of the heterogeneous graph, we define a variety of meta-paths to characterize the relatedness over different types of system entities from different views, which are summarized in Table 2. For example, ${M\!P}_4$ means that two processes are related if they are connected to files with the same attribute, while ${M\!P}_6$ indicates that two processed are related if they perform the same operation on the network.
\begin{table}[]
\caption{A summary of the meta-paths.}
\label{tab:two}
\resizebox{\columnwidth}{!}{
\begin{tabular}{ll}
\hline
\multicolumn{1}{c}{\textbf{ID}} & \multicolumn{1}{c}{\textbf{Meta-paths}}                          \\ \hline

${M\!P}_1$         & Process$\rightarrow$ op1 $\rightarrow$ Process                                                                                         \\
${M\!P}_2$         & Process$\rightarrow$contain$\rightarrow$Attribute$\rightarrow$contain$^{-1}$$\rightarrow$Process                                                                 \\
${M\!P}_3$        & Process$\rightarrow$op2$\rightarrow$File$\rightarrow$op2$^{-1}$$\rightarrow$Process                                                                              \\
${M\!P}_4$         & \begin{tabular}[c]{@{}l@{}}Process$\rightarrow$op2$\rightarrow$File$\rightarrow$contain$\rightarrow$Attribute$\rightarrow$contain$^{-1}$\\  $\rightarrow$File$\rightarrow$op2$^{-1}$$\rightarrow$Process\end{tabular} \\
${M\!P}_5$         & Process$\rightarrow$op5$\rightarrow$IPC$\rightarrow$op5$^{-1}$$\rightarrow$Process                                                                               \\
${M\!P}_6$         & Process$\rightarrow$op3$\rightarrow$Network$\rightarrow$op3$^{-1}$$\rightarrow$Process                                                                           \\
${M\!P}_7$         & Process$\rightarrow$op4$\rightarrow$Memory$\rightarrow$op4$^{-1}$$\rightarrow$Process                                                                            \\
${M\!P}_8$        & Process$\rightarrow$op6$\rightarrow$Socket$\rightarrow$op6$^{-1}$$\rightarrow$Process   \\\hline 
\end{tabular}}
\end{table}
After defining all the meta-paths, we apply HGAT, a heterogeneous graph embedding technique based on hierarchical attention\cite{wang2019heterogeneous}, to learn the d-dimensional vector for each node in the HPG. The advantage of HGAT over other heterogeneous graph embedding techniques (e.g., metapath2vec\cite{2017metapath2vec}, metagraph2vec\cite{zhang2018metagraph2vec}) is that it is able to learn the importance of different neighbors and meta-paths for a specific node. This characteristic is especially important for the cyber-attack detection task, since different system entities and behaviors usually have different sensitive and dangerous levels. We give an example to illustrate this phenomenon as follows.

\noindent \textbf{Example 1.} Given two meta-paths ${M\!P}_1$  (process$\rightarrow$ connect$\rightarrow$ network$\rightarrow$ connect$^{-1}\rightarrow$ process) and ${M\!P}_2$ (process$\rightarrow$ read $\rightarrow$ file$\rightarrow$ contain$\rightarrow$ sensitive data $\rightarrow$ contain$^{-1}\rightarrow$ file$\rightarrow$ read$^{-1}\rightarrow$ process), ${M\!P}_1$ represents the system behavior that the two processes connect to network, and ${M\!P}_2$ represents the two processes read sensitive data from files. Based on the experience in [11], stealing sensitive data from files is one of the most important reasons why the attackers launch a cyber-attack. Therefore, ${M\!P}_2$ plays a more important role than ${M\!P}_1$ for the cyber-attack detection task. Besides, when using ${M\!P}_2$, the entity type “sensitive data” should also be assigned a higher degree of importance than other entity types in ${M\!P}_2$ (e.g., “file”).

Specifically, we apply HGAT for heterogeneous graph embedding in two steps. In the first step, we generate sequences of system entities with various types from the heterogeneous graph by using the meta-paths to guide the random walker. Given a heterogeneous graph $G$ = ($V$, $E$) and a set of meta-paths $M\!P\!S$, the random walker works as follows. First, the random walker randomly chooses a meta-path ${M\!P}_i$ from ${M\!P}_S$ (in the form of $A_1\rightarrow R_1\rightarrow …\rightarrow A_t\rightarrow R_t\rightarrow A_{t+1}\rightarrow …\rightarrow R_{L-1}\rightarrow A_L$). Second, the random walker travels on the heterogeneous graph in accordance with ${M\!P}_i$, and the transition probability from node $v_j$ to $v_{j+1}$ at step $j$ is defined in Equation \ref{eq:one}, where $\varphi$($v_j$) = $A_t$ and $N_{t+1}(v_j)$ is the set of neighbors of node $v_j$ with entity type $A_{t+1}$. Finally, the random walker randomly picks another meta-path ${M\!P}_{i+1}$ starting with $A_L$, and repeats the transition process.

\begin{tiny}
\begin{equation}
 \label{eq:one}
 p({v_{j + 1}}|{v_j},M\!{P_i})=
 \begin{cases}
  \frac{1}{|{N_{t + 1}}({v_j})|} & ({v_j},{v_{j + 1}}) \in E, \phi ({v_{j + 1}}) = {A_{t + 1}} \\
  \displaystyle 0 & ({v_j},{v_{j + 1}}) \in E,{\kern 1pt} {\kern 1pt} {\kern 1pt} \phi ({v_{j + 1}}) \ne {A_{t + 1}} \\
  \displaystyle 0 & ({v_j},{v_{j + 1}}) \notin E\\
  
 \end{cases}
\end{equation}
\end{tiny}

In the second step, we feed the generated system entity sequences to the HGAT model to learn the system entity embeddings. HGAT is a GNN based on a hierarchical attention structure (i.e., a node-level attention and a semantic-level attention)\cite{wang2019heterogeneous}. Specifically, for a given system entity vi and a meta-path ${M\!P}_j$, the node-level attention learns the importance of each neighbor of vi in the generated system entity sequence from ${M\!P}_j$, and aggregates these meaningful neighbors to form a candidate embedding of $v_i$ (denoted as $ z_i^j$ ). Then, the semantic-level attention learns the importance of different meta-paths for $v_i$, and fuses all the candidate embeddings (i.e.,$z_i^1,z_i^2,z_i^{M\!P\!S}$  ) to obtain the final embedding of $v_i$. Note that HGAT should be trained with a downstream task. Here, we utilize system entity classification (i.e., classifying a “process” as benign or malicious) as the downstream task.

\subsection{Local Graph Sampling and Cyber-Attack Detection}

After obtaining all the system entity embeddings, the simplest strategy of detecting cyber-attacks is to train a classifier to classify each “process” as benign process or cyber-attack process, by directly taking the $d$-dimensional embedding vectors of system entities as features. However, this simple strategy has the following problems. First, cyber-attacks are usually stealthy, so it is very difficult to detect them by only considering one system entity. Second, the cyber-attack detection system has to continuously deal with new incoming system entities, while it is infeasible to frequently rerun the heterogeneous graph embedding procedure.

To address these problems, we propose a local graph sampling based detection strategy. Given a new incoming system entity $v_n$ and the system entities interact with $v_n$ (denoted as $N\!V\!S$), we firstly link each system entity in $N\!V\!S$ to the existing heterogeneous graph (the combined graph is denoted as $CG$). Note that the entity type “attribute” can guarantee that every system entity will be linked to the existing heterogeneous graph. Then, we sample a local graph from CG for $N\!V\!S$ based on the k-order sub-graph sampling defined as follows.

\noindent \textbf{Definition 4 ($k$-order sub-graph sampling).} For a system entity $v_i$, the k-order sub-graph sampling of vi on $G$ = ($V$, $E$) (denoted as $SG(v_i, G)$) is defined in Equation \ref{eq:two}. Based on the definition, every system entity can form a local graph based on the k-order sub-graph sampling. For a set of system entities $V\!S$, the k-order sub-graph sampling of $V\!S$ on $G$ = ($V$, $E$) (denoted as $SG(VS, G)$) is defined in Equation \ref{eq:three}.

\begin{scriptsize}
\begin{equation}
 \label{eq:two}
 S{G^{(k)}}({v_i},G) =
 \begin{cases}
  \displaystyle{\{ {v_j}|({v_i},{v_j}) \in E\} }&{k = 1} \\
  \displaystyle {\{ S{G^{(1)}}({v_z},G)|{v_z} \in S{G^{(k - 1)}}({v_i},G)\} }&{k > 1} \\
 \end{cases}
\end{equation}
\end{scriptsize}

\begin{equation}
 \label{eq:three}
 S{G^{(k)}}(VS,G) =\displaystyle \bigcup\nolimits_{{v_i} \in VS} {S{G^{(k)}}({v_i},G)} \\
\end{equation}

\begin{figure}
    \centering
    \includegraphics[width=0.9\columnwidth]{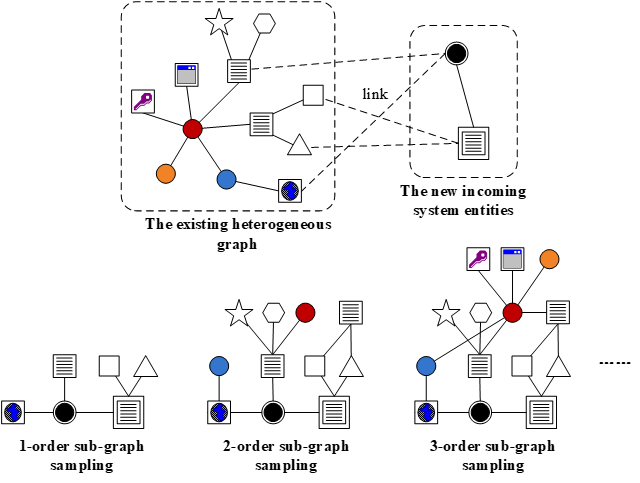}
    \caption{An example of the local graph sampling strategy.}
    \label{fig:four}
\end{figure}

The local graph could grow too large without constraint. For example, if the local graph contained an internal node $v_i$ with type “common file”, it would connect to too many nodes with type “file”. To prevent such situation, we put a threshold on the degree of nodes in the local graph. Specifically, we firstly categorized the seven major entity types into two groups, i.e., instantiable entity type (i.e., process and file) and conceptual entity type (i.e., socket, IPC, memory, network, and attribute). Obviously, the number of nodes with conceptual entity type is limited, and thus we only consider the number of nodes with instantiable entity type. For each node in the local graph, the number of its connected nodes with instantiable entity type should be no more than $\lambda$. Figure \ref{fig:four} shows an example of the local graph sampling strategy. The purpose of local graph sampling is twofold. First, rather than utilizing the information of only one system entity, we will aggregate the context information from all system entities in the local graph for cyber-attack detection. Second, for a new incoming unknown system entity, we will infer its embedding based on the information transferred from the neighbors in the local graph, with no need to rerun the heterogeneous graph embedding process.

To achieve the above purposes, we perform cyber-attack detection as a classification task for each node with type “process” in the local graph. Specifically, we apply R-GCN (Relational Graph Convolutional Network)\cite{schlichtkrull2018modeling} for the node classification task. R-GCN is an extension of GCN to relational graphs. It accumulates information from neighbors to a central node (denoted by $v_i$) based on Equation \ref{eq:four}, where $h_i^{(l)}$ is the hidden state of node $v_i$ in the $l$-th R-GCN layer, $N_i^r$ is the set of indices of neighbors of $v_i$ with edge type $r$, $W_0^{(l)}$ and $W_r^{(l)}$ are learnable parameters. Note that more R-GCN layers (i.e., a larger value of $l$) indicate the model can aggregate information from more remote nodes. 

\begin{equation}
 \label{eq:four}
 h_i^{(l + 1)} = \sum\limits_{r \in R} {\sum\limits_{j \in N_i^r} {\frac{1}{{|N_i^r|}}W_r^{(l)}h_j^{(l)} + W_0^{(l)}h_i^{(l)}} } \\
\end{equation}

R-GCN can be viewed as an encoder of the local graph. By stacking multiple R-GCN layers on the local graph, higher order of context information in the local graph could be learnt. Then, we simply add a softmax activation on the output hidden state of each node of the last layer by minimizing the cross-entropy loss on all labelled nodes to train the node classifier.

\section{Experiment}
\label{sec:experiment}

\subsection{Experiment Setup}

\begin{table*}[]

\caption{The details of the datasets.}
\label{tab:three}
\resizebox{\textwidth}{!}{
\begin{tabular}{llllllll}
\hline
\textbf{Dataset}       & \textbf{\begin{tabular}[c]{@{}l@{}}Duration\\    (hh:mm:ss)\end{tabular}} & \textbf{File read} & \textbf{File write} & \textbf{Process/Thread} & \textbf{Network} & \textbf{Others} & \textbf{\#Attack} \\ \hline
\textbf{Lab Dataset}   & 10:41:33                                                                  & 45.11\%            & 18.41\%             & 1.94\%                  & 14.47\%          & 20.07\%         & 34                \\
\textbf{DARPA Dataset} & 24:32:15                                                                  & 45.10\%            & 27.78\%             & 0.91\%                  & 9.04\%           & 17.17\%         & 7     \\\hline           
\end{tabular}}
\end{table*}

\subsubsection{Dataset}
We evaluate our method using the following two different datasets with multiple types of cyber-attacks.

\noindent \textbf{Lab Dataset}: We simulated cyber-attacks on a real host and collected datasets with ground truth for evaluation. The details of this dataset are shown in the first row of Table \ref{tab:three}. The dataset was collected from Linux. The column “Duration” refers to the length of time that the collector was running on the target host, covering both normal activities (e.g., watching online videos, website browsing, document editing, etc.) and attack-related activities. The next several columns provide a breakdown of events into different types of operations. The column “\#Attack” refers to the number of cyber-attack campaigns. Note that a cyber-attack campaign could involve a lot of malicious activities and malicious processes.

There are mainly three types of cyber-attacks performed during the collection of this dataset. The first type is “attack by webshell”, which uses webshell scripts to escalate and maintain persistent access to the target system by exploiting web vulnerabilities. The second type is “attack by RAT (Remote Access Trojan)”, which implants Trojans to the target system through phishing attacks and then uses the Trojans to steal sensitive information. The third type is “attack by LotL (Living off the Land)”, which runs malicious shellcode directly in memory.

Next, we used SPADE\cite{gehani2012spade}, a provenance graph construction toolkit, to convert the collected data into a provenance dataset in the form of provenance graphs. Finally, the dataset contains 19659 processes, 10073 files, and 2700912 system events.

\noindent \textbf{DARPA Dataset}: This dataset was collected from three hosts installed Windows 10 during a cyber-attack program carried out by a read team as part of the DARPA Transparent Computing program. The details of this datasets are shown in the second row of Table \ref{tab:three}.

There are mainly three types of cyber-attacks performed during the collection of this dataset, i.e., sensitive information gathering and exfiltration, in-memory attack by exploiting FireFox vulnerabilities, and malicious file downloading and execution. It can be seen that the strategies of cyber-attacks in the DARPA dataset are quite different from those in the Lab dataset. However, all these strategies contain one or more invariant stages (e.g., deploying malicious code, stealing sensitive information, communicating with C\&C server).

In addition, while the read team was attacking the target host, benign background activities were also being performed on the host. In general, nearly 99.9\% of the system events are related to benign activities. Finally, after the provenance graph construction and compaction, the dataset contains 53084 processes, 219706 files, and 26751468 system events. 

We conducted the experiments on a server with an Intel Xeon E5-2680 v4 CPU (with 14 × 2.4 GHz cores), 128GB of memory, and 4 × GTX 2080Ti GPUs running on Ubuntu 20.04.
\subsubsection{Evaluation Strategy}
First, we introduce the two evaluation strategies used in our experiments as follows.

1) In-Sample: This evaluation strategy is used to evaluate the cyber-attack detection performance on the existing system entities. Specifically, it builds the heterogeneous graph on the whole dataset, which is then divided into training set and testing set. To maintain the original proportion of malicious / benign processes in both training and testing sets, we use stratified sampling to choose two-thirds of the processes with malicious label and two-thirds of the processes with benign label to generate the training set, and use the rest as testing set.

2) Out-Sample: This evaluation strategy is used to evaluate the cyber-attack detection performance on new incoming processes. Specifically, it adopts stratified sampling to divide the whole dataset into training set (two-thirds of the processes) and testing set (one-third of the processes). The training set is used for heterogeneous graph construction, heterogeneous graph embedding, and cyber-attack detector training. The testing set is used as new incoming processes to evaluate the cyber-attack detector in an online way.

Second, we use ACC, Precision, Recall, and Macro-F1 as the performance metrics. Here, ACC refers to accuracy. Precision and Recall are only calculated for detecting malicious samples. Note that Precision is inversely proportional to the false alarm rate. Macro-F1 is used with the consideration of the imbalanced malicious / benign samples.
\subsubsection{Training Strategy}
The dataset is highly class imbalanced, i.e., the majority of training samples are benign processes. Thus, standard training strategy tends to be overwhelmed by the benign samples and ignore the malicious ones. Aiming at this situation, we adopt a cost-sensitive version of the cross entropy loss function. Specifically, we assign a weight $\delta$ to each sample, i.e., $\delta$ = 1 for each malicious sample and $\delta$ = 0.1 for each benign sample.
\subsubsection{Parameter Settings}
There are three key parameters in our cyber-attack detection method, i.e., the dimension of the node embeddings $d$, the threshold of the node degree in the local graph $\lambda$, and the order of the sub-graph sampling $k$. Through several grid search experiments, we finally set $d$ = 32, $\lambda$ = 10, and $k$ = 8.
\subsection{Ablation Experiment}
\begin{figure}
    \centering
    \includegraphics[width=0.9\columnwidth]{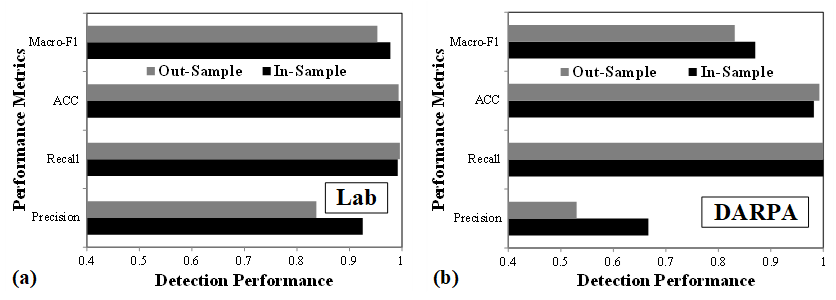}
    \caption{The evaluation of our method for classifying new incoming unknown processes: (a) the experiment results on Lab Dataset; (b) the experiment results on DARPA Dataset.}
    \label{fig:five}
\end{figure}

\begin{table*}[]
\caption{The computation overhead of cyber-attack detection.}
\label{tab:four}
\resizebox{\textwidth}{!}{
\begin{tabular}{lllllll}
\hline
\multirow{2}{*}{}       & \multicolumn{5}{c}{\textbf{Whole-Graph}}                                                                                                                                                                       & \multicolumn{1}{c}{\multirow{2}{*}{\textbf{Local-Graph}}} \\ \cline{2-6}
                        & \multicolumn{1}{c}{\textbf{50 nodes}} & \multicolumn{1}{c}{\textbf{500 nodes}} & \multicolumn{1}{c}{\textbf{5000 nodes}} & \multicolumn{1}{c}{\textbf{50000 nodes}} & \multicolumn{1}{c}{\textbf{90000 nodes}} & \multicolumn{1}{c}{}                                      \\ \hline
\textbf{Memory Usage}   & 0.821GB                               & 0.981GB                                & 1.439GB                                 & 6.361GB                                  & 10.263GB                                 & 0.845GB                                                   \\
\textbf{Execution Time} & 0.1558s                               & 0.1602s                                & 0.1669s                                 & 0.6568s                                  & 2.1332s                                  & 0.1598s        \\\hline                                          
\end{tabular}}
\end{table*}
In the first experiment, we try to evaluate the capability of our method for classifying new incoming unknown processes. Specifically, we compare the detection performance of our method under In-Sample and Out-Sample evaluation strategies. The experiment results are shown in Figure \ref{fig:five}. Out-Sample has a lower Precision and almost the same Recall as compared with In-Sample, which means that the false alarm rate increases to a certain extent on unknown processes. This result is reasonable that the detection performance would be affected without a full view of the relations between the target process and other system entities. However, Out-Sample has almost the same ACC and a slight lower Macro-F1 as compared with In-Sample. It demonstrates that our method could still guarantee a satisfied overall detection performance on unknown processes.

In the second experiment, we try to verify the effectiveness of the local graph sampling component of our method. To be specific, we evaluate the detection performance and computation complexity of the following two variants. In this experiment, In-Sample evaluation strategy is utilized.

1) Whole-Graph: It treats the whole heterogeneous graph as the local graph.

2) Local-Graph: It samples the 8-order sub-graph as the local graph.

\begin{figure}
    \centering
    \includegraphics[width=0.9\columnwidth]{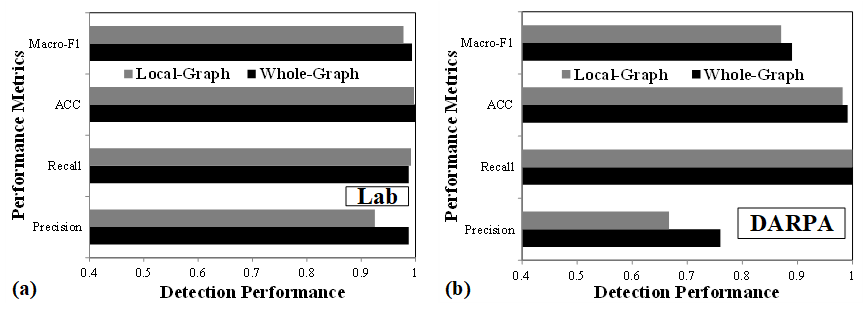}
    \caption{The evaluation of the local graph sampling component: (a) the experiment results on Lab Dataset; (b) the experiment results on DARPA Dataset.}
    \label{fig:six}
\end{figure}

The detection performance is shown in Figure \ref{fig:six}. From the figure we can see that Whole-Graph outperforms Local-Graph to a slight degree. It implies that some cyber-attacks are very stealthy, i.e., they do not perform malicious activities immediately after entering the system, but often conceal themselves for a long time and trigger the attacks by an alternative system entity. It would result in a long distance between the malicious process and the manipulated system entity performing the malicious activities in the heterogeneous graph. However, processing the whole heterogeneous graph would lead to extremely high computation overhead, especially the memory usage. To verify this assumption, we test the memory usage and the execution time of cyber-attack detection on the whole heterogeneous graph. We simulate very large heterogeneous graphs by randomly inserting fake nodes and edges (each node has at most 100 edges). Table \ref{tab:four} shows the test results, where “Execution Time” includes the time for performing convolution operation on the graph and classifying a single node. As shown in Table \ref{tab:four}, as the number of nodes increases, the execution time becomes slightly longer, but the memory usage exhibits a significant growth. Note that the execution time and memory usage do not increase linearly, since there exists common computation overhead for building and storing the heterogeneous graphs.
In the third experiment, we try to verify the effectiveness of the heterogeneous graph embedding component of our method. Specifically, we test the detection performance of combinations of different meta-paths (i.e., MPC1 to MPC6 as follows) on the Lab Dataset. In this experiment, In-Sample evaluation strategy is utilized.

1) MPC1: It includes the meta-paths that only contain process and file (i.e., MP1 and MP3).

2) MPC2: It includes the meta-paths that only contain process, file, network, and socket (i.e., MP1, MP3, MP6, and MP8).

3) MPC3: It includes the meta-paths that only contain process, file, IPC, and memory (i.e., MP1, MP3, MP5, and MP7).

4) MPC4: It includes the meta-paths that only contain process, file, and file attribute (i.e., MP1, MP3, and MP4).

5) MPC5: It includes the meta-paths that only contain process, file, and process attribute (i.e., MP1, MP2, and MP3).

6) MPC6: It includes all meta-paths.

The experiment results are shown in Table \ref{tab:five}. First, MPC6 has the best overall detection performance. It demonstrates that all the meta-paths contribute to the cyber-attack detection task, which also implies that cyber-attacks are highly complex and it is impossible to detect all cyber-attacks by only considering certain types of system entities and system events. Second, MPC4 and MPC5 outperform MPC1. It shows that process and file attributes can provide richer information to improve the detection performance. Third, MPC4 outperforms all other methods, except for MPC6. It means that file attributes are the most important information for cyber-attack detection. This result is in consistent with the existing experience that stealing sensitive data from files or performing sensitive operations on files is one of the most important reasons for cyber-attacks [11]. Fourth, MPC2 outperforms MPC3 and MPC5. It implies that accessing the network is a more important characteristic of cyber-attacks than manipulating the processes and memory.

\begin{table}[]
\caption{The evaluation of different meta-path combinations.}
\label{tab:five}
\resizebox{\columnwidth}{!}{
\begin{tabular}{lllll}
\hline
     & \textbf{Precision} & \textbf{Recall} & \textbf{ACC}    & \textbf{Macro-F1} \\ \hline
MPC1 & 0.8514             & 0.9916          & 0.9940          & 0.9565            \\
MPC2 & 0.8902             & 0.9916          & 0.9957          & 0.9679            \\
MPC3 & 0.8708             & 0.9958          & 0.9950          & 0.9633            \\
MPC4 & 0.9046             & \textbf{1.0000} & 0.9965          & 0.9740            \\
MPC5 & 0.8551             & 0.9958          & 0.9943          & 0.9586            \\
MPC6 & 0.9252             & 0.9916          & \textbf{0.9971} & \textbf{0.9779} 
\\\hline
\end{tabular}}
\end{table}\
\subsection{Comparison Experiment}
In the first experiment, we evaluate the general detection performance of our method (abbreviated as OUR) by comparing it with the following four baselines. All the baselines have considered the class imbalance problem and assign different weights to samples (10:1 for malicious samples and benign samples) during the training process. In this experiment, In-Sample evaluation strategy is utilized.

1) SVM: It refers to the process classification model, which uses the vector space model to create the feature vector (each entry represents the number of nodes of a certain type in the first-order neighbors of the target process node in the heterogeneous graph) and the SVM as classifier.

2) GAT: It refers to a homogeneous graph neural network\cite{velivckovic2017graph}. Specifically, it first constructs a homogeneous graph of the same topology with heterogeneous graph, but ignores the types of nodes, and then performs node embedding and classification based on the GAT model.

3) HGAT: It refers to a heterogeneous graph neural network\cite{wang2019heterogeneous}. Specifically, it first constructs the heterogeneous graph, and then performs node embedding and classification based on the HGAT model, without the local graph sampling and the R-GCN steps. It can be viewed as a variant of our method by only looking at one process.

4) CONAN: It refers to a state-of-the-art rule based cyber-attack detection model\cite{0CONAN}. It manually defines a large number of rules (called atomic suspicious indicators) from four aspects (i.e., code source, behavior, feature and network) to support the cyber-attack detection task.

The experiment results are shown in Table \ref{tab:six}. First, GAT is completely unable to detect cyber-attacks, because that GAT only considers the interactions between system entities but ignores all the semantics (e.g., the type of system entities, the context of system events, etc.), which are essential for detecting malicious activities. Second, SVM also has a poor detection performance, because that SVM only considers the types of system entities adjacent to the target process but ignores the downstream activities of these system entities. In practice, the attacker would always conceal themselves for a certain period of time before performing the malicious activities. Third, our method outperforms HGAT. It again shows that it is difficult to detect cyber-attacks by only looking at one system entity. Fourth, CONAN has a slight advantage over our method on the Lab Dataset, while our method outperforms CONAN on the DARPA Dataset. CONAN is a rule-based cyber-attack detection system manually designed based on an in-depth analysis on the attack samples. Although the rules could perfectly adapt to the analysed attack samples, they cannot handle the changing patterns and thus exhibit a far lower performance in a different macro environment. On the other hand, our method is totally data-driven and does not rely on in-depth domain knowledge. It can be re-trained periodically or when new attack samples are collected, to automatically adapt to the change of macro environment. Thus, our method has significant advantage over CONAN on generalization ability.

\begin{table}[]
\caption{The comparison of different in-sample detection methods.}
\label{tab:six}
\resizebox{\columnwidth}{!}{
\begin{tabular}{lllll}
\hline
               & \textbf{Precision}         & \textbf{Recall} & \textbf{ACC}    & \textbf{Macro-F1} \\ \hline
\multicolumn{2}{l}{\textbf{Lab Dataset:}}   &                 &                 &                   \\
SVM            & \textbf{1.0000}            & 0.0457          & 0.9641          & 0.5041            \\
GAT            & 0.0000                     & 0.0000          & 0.9670          & 0.4916            \\
HGAT           & 0.7102                     & 0.1093          & 0.9663          & 0.5124            \\
CONAN          & 0.9758                     & \textbf{0.9981} & \textbf{0.9983} & \textbf{0.9915}   \\
OUR            & 0.9252                     & 0.9916          & 0.9971          & 0.9779            \\
\multicolumn{2}{l}{\textbf{DARPA Dataset:}} &                 &                 &                   \\
SVM            & 1.0000                     & 0.1253          & 0.9237          & 0.5709            \\
GAT            & 0.0000                     & 0.0000          & 0.9715          & 0.4925            \\
HGAT           & \textbf{1.0000}            & 0.2491          & 0.9732          & 0.6932            \\
CONAN          & 0.5895                     & 1.0000          & 0.9759          & 0.8395            \\
OUR            & 0.6667                     & \textbf{1.0000} & \textbf{0.9820} & \textbf{0.8703}  
\\\hline
\end{tabular}}
\end{table}
In the second experiment, we focus on the detection performance on unknown processes by comparing our method with the following two baselines. In this experiment, Out-Sample evaluation strategy is utilized.

1) NeighborAvg: It is a variant of our method. Specifically, a new incoming unknown process is linked to the existing heterogeneous graph and represented by averaging the embed-dings of its neighbors.

2) LabelProp: It is a variant of our method. Specifically, a new incoming unknown process is first linked to the existing heterogeneous graph, and then Label Propagation (a semi-supervised learning algorithm)\cite{xiaojin2002learning} is used to propagate labels of the existing nodes to the unknown process.

The experiment results are shown in Table \ref{tab:seven}. First, NeighborAvg has a poor detection performance. It is because that NeighborAvg only considers the first-order neighbors and cannot adapt to the stealth of cyber-attacks, which would often result in a long distance between the malicious process and the system entity performing the malicious activities in the heterogeneous graph. Therefore, LabelProp has a far better performance than NeighborAvg, since it considers higher order correlations by propagating labels to a long distance. Second, our method outperforms LabelProp. It is because that LabelProp only transfers malicious labels without capturing the patterns of malicious activities. There are usually no malicious system entities in a certain range of the malicious process, especially only a local graph is sampled for detection.
\begin{table}[]
\caption{The comparison of different out-sample detection methods.}
\label{tab:seven}
\resizebox{\columnwidth}{!}{
\begin{tabular}{lllll}
\hline
                   & \textbf{Precision}       & \textbf{Recall} & \textbf{ACC}    & \textbf{Macro-F1} \\ \hline
\multicolumn{2}{l}{\textbf{Lab Dataset:}}     &                 &                 &                   \\
NeighborAvg        & 0.9151                   & 0.1416          & 0.9151          & 0.5634            \\
LabelProp          & \textbf{0.9620}          & 0.5067          & 0.9678          & 0.8234            \\
OUR                & 0.8369                   & \textbf{0.9958} & \textbf{0.9935} & \textbf{0.9530}   \\
\multicolumn{2}{l}{\textbf{DARPA   Dataset:}} &                 &                 &                   \\
NeighborAvg        & 0.6679                   & 0.1258          & 0.8839          & 0.5739            \\
LabelProp          & \textbf{0.7500}          & 0.3312          & 0.9350          & 0.6971            \\
OUR                & 0.5300                   & \textbf{1.0000} & \textbf{0.9924} & \textbf{0.8314}
\\\hline
\end{tabular}}
\end{table}
\subsubsection{Case Study}
In this section, we provide a real attack case to illustrate how our method works. Given the provenance graph of an attack scenario in Figure \ref{fig:one}, where the diamonds represent network, the ovals represent processes, and the rectangles represents files. In this case, the attacker first penetrates the Apache server and leaves a backdoor. When the attacker triggers the backdoor, it creates a process (i.e., “Webshell”). Then, “Webshell” forks several processes to perform sensitive commands (e.g., “netstat” to check the detailed network information) and writes the collected information to a file (i.e., “mailer.log”). Finally, “Webshell” executes a process (i.e., “mailman”) to collect the sensitive information from “mailer.log” and another sensitive file (i.e., “shadow”), and send to a remote server.

\begin{figure}
    \centering
    \includegraphics[width=0.9\columnwidth]{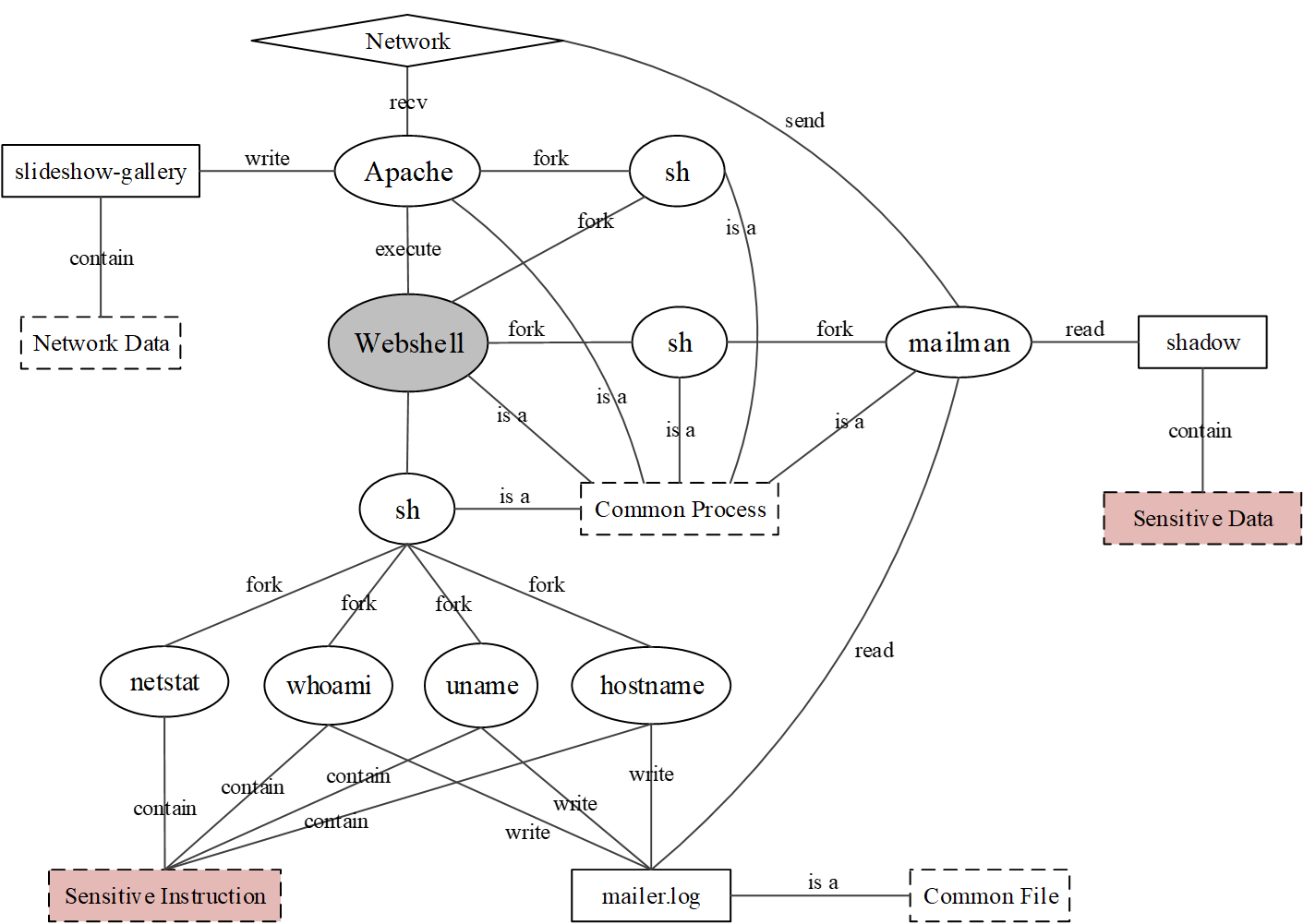}
    \caption{The heterogeneous graph of a real attack case.}
    \label{fig:seven}
\end{figure}

The heterogeneous graph transformed from Figure 1 is partially shown in Figure \ref{fig:seven}, where the rectangles with dashed border represent attributes. First, according to\cite{0CONAN}, the final purpose of cyber-attacks is stealing sensitive information or causing damage, and thus they usually share some common features (e.g., performing sensitive instructions, reading sensitive files, etc.). These common features can be learnt by the meta-path based heterogeneous graph embedding. Second, there is usually a certain distance between the original malicious process and the system entities performing the malicious activities (e.g., “Webshell” and “netstat”, “Webshell” and “reading from shadow”). Therefore, only looking at one system entity is unable to detect such kind of cyber-attacks, while our method can model a chain of system entities based on local graph sampling and convolution operations. For example, take “Webshell” as the process to be classified, if we sample a 3-order sub-graph from it as the local graph, the “Sensitive Instruction” feature can be propagated to it through convolution operations. If we sample a 4-order sub-graph from it as the local graph, the “Sensitive Data” feature can also be propagated to it. These features can potentially help the model to decide “Webshell”is a malicious process.
\section{Conclusions and Future Work}
\label{sec:conclusion}
In this paper, we investigate the cyber-attack detection problem based on provenance data. We propose an intelligent cyber-attack detection system based on deep learning technique. Specifically, it adopts a heterogeneous provenance graph to model all system entities and events, and learns a low-dimensional vector to represent each system entity in a scalable way. Then, it reconstructs the attack scene by sampling a small sub-graph from the provenance graph and detects cyber-attacks on this sub-graph. Based on the above designs, our method could effectively detect cyber-attacks with the characteristics of persistence, stealth, and diversity. Through a series of experiments based on provenance datasets containing real cyber-attacks, we demonstrate that our method outperforms other current machine learning and deep learning based models, and has competitive performance as compared with state-of-the-art rule-based cyber-attack detection systems designed by excessive domain knowledge.

{\small
\bibliographystyle{ieee_fullname}
\bibliography{egbib}
}

\end{document}